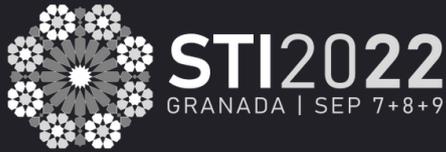

*Special track*

**STI 2022 Conference Proceedings**
*Proceedings of the 26th International Conference on Science and Technology Indicators*

All papers published in this conference proceedings have been peer reviewed through a peer review process administered by the proceedings Editors. Reviews were conducted by expert referees to the professional and scientific standards expected of a conference proceedings.

**Proceeding Editors**

Nicolas Robinson-Garcia
Daniel Torres-Salinas
Wenceslao Arroyo-Machado

26th International Conference on Science and Technology Indicators | **STI 20**22

## "From Global Indicators to Local Applications"

7–9 September 2022 | Granada, Spain

**#STI22GRX**

# Persistent Identification and Interlinking of FAIR Scholarly Knowledge[1]

Muhammad Haris[*], Markus Stocker[**] and Sören Auer[**]

[*]*haris@l3s.de*
L3S Research Center, Leibniz University Hannover, Hannover 30167 (Germany)

[**]*markus.stocker@tib.eu; auer@tib.eu*
TIB---Leibniz Information Centre for Science and Technology, Leibniz University Hannover, Hannover 30167 (Germany)

**Abstract**

We leverage the Open Research Knowledge Graph - a scholarly infrastructure that supports the creation, curation, and reuse of structured, semantic scholarly knowledge - and present an approach for persistent identification of FAIR scholarly knowledge. We propose a DOI-based persistent identification of ORKG Papers, which are machine-actionable descriptions of the essential information published in scholarly articles. This enables the citability of FAIR scholarly knowledge and its discovery in global scholarly communication infrastructures (e.g., DataCite, OpenAIRE, and ORCID). While publishing, the state of the ORKG Paper is saved and cannot be further edited. To allow for updating published versions, ORKG supports creating new versions, which are linked in provenance chains. We demonstrate the linking of FAIR scholarly knowledge with digital artefacts (articles), agents (researchers) and other objects (organizations). We persistently identify FAIR scholarly knowledge (namely, ORKG Papers and ORKG Comparisons as collections of ORKG Papers) by leveraging DataCite services. Given the existing interoperability between DataCite, Crossref, OpenAIRE and ORCID, sharing metadata with DataCite ensures global findability of FAIR scholarly knowledge in scholarly communication infrastructures.

## Introduction

The number of scholarly articles is rapidly increasing every year (Jinha, 2010). Traditional scientific publishing is document-based (i.e., PDF documents); therefore, it is tedious for researchers to filter the articles that meet their information needs. For example, it is not possible to retrieve all articles reported SARS-CoV-2 R0 basic reproductive numbers because the scholarly content cannot be efficiently extracted by machines from PDF documents. The Open Research Knowledge Graph (ORKG) (Jaradeh, et al., 2019) addresses this problem by representing scholarly knowledge in a machine actionable, structured, and semantic manner, and providing services for comparing scholarly knowledge (Oelen et al., 2020) in tabular and other visual forms (Wiens et al., 2020) as well as other types of services for efficient reuse of scholarly knowledge. ORKG supports describing scholarly articles in the form of research contributions, whereby a contribution represents the results obtained by means of some materials and methods and addressing a research problem. The scholarly knowledge created in ORKG strongly adheres to the FAIR principles (Wilkinson et al., 2016), but this knowledge is currently not discoverable in global scholarly communication infrastructures. Towards this goal, we propose to leverage ORKG and persistently identify its FAIR scholarly knowledge (Papers). This work builds and extends our earlier work published in Haris et al. (2021).

---

[1] This work was co-funded by the European Research Council for the project ScienceGRAPH (Grant agreement ID: 819536) and TIB--Leibniz Information Centre for Science and Technology.

Our contributions are as follows:

1-      Persistently identify FAIR scholarly knowledge (specifically, ORKG Papers) by leveraging DataCite services, thereby ensuring the broad findability of machine actionable scholarly knowledge in global scholarly communication infrastructures (DataCite, OpenAIRE and ORCID).

2-      Support updating published ORKG Papers as separate versions. These versions are linked in provenance chains that track changes over time.

We address the following research question: How can FAIR scholarly knowledge be made findable in global scholarly communication infrastructures?

**Related Work**

Persistent Identifiers (PIDs) play an important role in persistently identifying research articles, datasets, software, and other digital artefacts, as well as physical objects like samples (Paskin, 2010). PIDs are an essential component of scholarly communication infrastructures, as they are less likely to result in broken references than traditional hyperlinks because they keep the identifier separate from location of a digital resource on the Web. This ensures the persistent reference of the scholarly record.

Several organizations provide services for persistently identifying scholarly objects. Crossref and DataCite provide DOI-based persistent identification of scholarly articles and datasets, respectively, whereas ORCID (Haak, Fenner, Paglione, Pentz, & Ratner, 2012) enables the persistent identification of researchers. There are several emerging identification schemes, including the Research Organization Registry (ROR) for organizations, the International Geo Sample Number (IGSN) for samples, as well as approaches for scientific instruments (Stocker et al., 2020). The structured comparison of various persistent identifier systems can be found in Auer & Stocker (2021).

Persistent identifiers have associated metadata, which is managed separately from the identified artefact. This metadata information enables artefact findability and accessibility, as well as metadata linking and sharing within scholarly infrastructures (Meadows, Haak, and Brown, 2019). Wilkinson et al. (2016) emphasize the need for permanent identification of research artefacts in relation to FAIR data principles. Persistent identifiers are also used for a variety of entity types, and their implementation is seen as crucial for scholarly communication infrastructures. Richards et al. (2011) addressed persistent identification of datasets; Stocker et al. (2020) of instruments; Franken et al. (2022) of conferences; Farjana et al. (2016) of geometric and topological entities. Bellini et al. (2012) developed an ontologybased technique for enriching metadata in PID datasets that facilitates interoperability amongst PID systems.

Figure 1: Conceptual model for persistently identifying FAIR scholarly knowledge and ensuring its findability in global scholarly communication infrastructures.

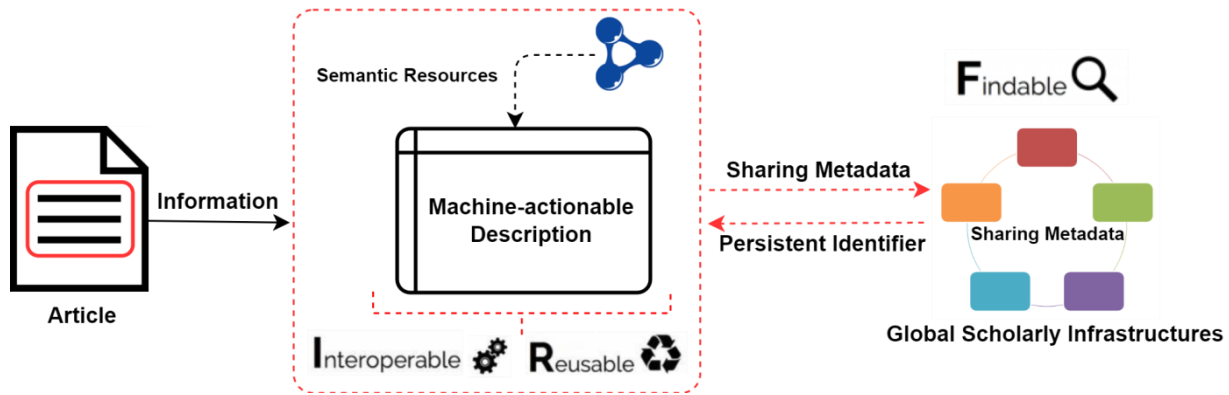

## Conceptual Model

In this section, we propose a conceptual model for the persistent identification of FAIR scholarly knowledge, i.e., machine-actionable ORKG Papers. The model is shown in Figure 1 and comprises the following key aspects:

1.       Essential knowledge is extracted from scholarly articles and represented in machine actionable form by semantifying knowledge through linking to third-party semantic resources, e.g., ontologies. Semantically rich, machine-actionable scholarly knowledge is persistently identified and metadata about published scholarly knowledge (ORKG Papers) is shared with scholarly infrastructures. The persistent identification of machine-actionable scholarly knowledge ensures broad findability and further improves reusability through the use of unambiguous identifiers with associated machine actionable metadata.

2.       Support updating published ORKG Papers as individual stable versions as well as the persistent identification of these versions to convey the latest updates. With persistent identification, provenance is also tracked to show what updates have been made to the content in each version over time.

## Approach

In this section, we present the implementation of the proposed conceptual model. Aligned with the model, we cover two main aspects. First, persistently identifying scholarly knowledge (specifically, ORKG Papers) by means of DOI. Second, support updating already persistently identified ORKG Papers.

### Persistent Identification of ORKG Papers

ORKG Papers are machine actionable descriptions of essential information published in scholarly articles. These Papers can be persistently identified by leveraging DataCite and creating, accordingly, metadata including title, description, research field and creators of the ORKG Paper.

Figure 2 shows how essential scholarly knowledge is extracted from a scholarly article and represented in ORKG in the form of Research Contributions. The scholarly knowledge curated in ORKG strongly complies with FAIR data principles. This machine-actionable scholarly knowledge is persistently identified by leveraging DataCite services and publishing metadata through its REST API[2] by following its metadata schema[3]. Moreover, the published versions can be updated as separate versions and each version can be persistently identified by sharing metadata about it. The DOI-based persistent identification of ORKG scholarly knowledge enables its findability in global scholarly communication infrastructures.

Figure 2: Essential scholarly knowledge is extracted from scholarly articles and represented in ORKG in the form of research contributions (results addressing some research problem and obtained by means of some materials and methods). Metadata about the ORKG Paper is shared with DataCite to persistently identify the FAIR scholarly knowledge. This ensures the broad findability of machine actionable scholarly knowledge in global scholarly communication infrastructures (DataCite, Crossref, OpenAIRE, ORCID, and PID Graph).

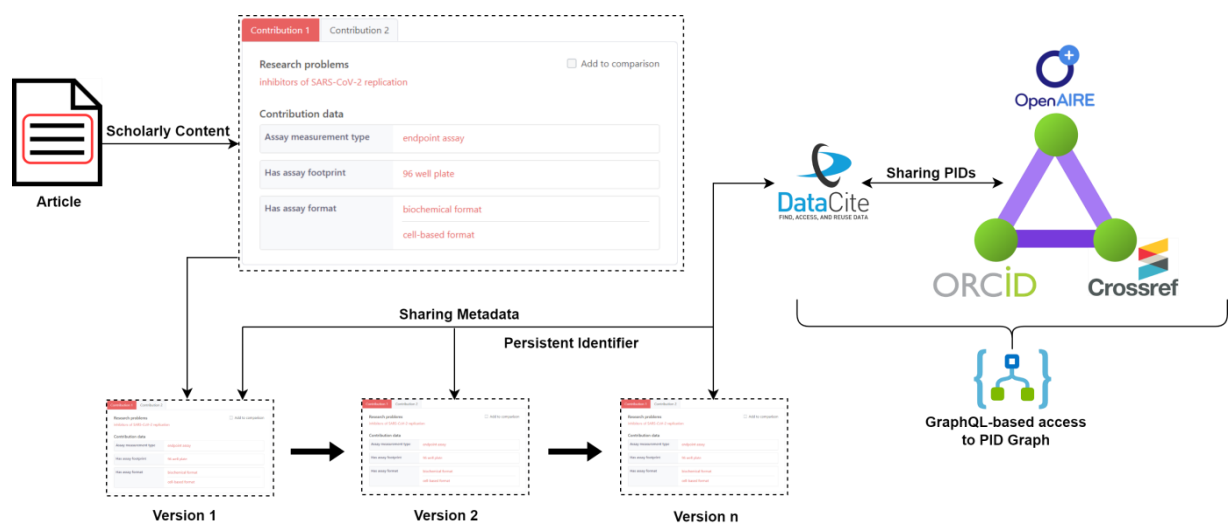

While publishing the ORKG Paper, it is ensured that the metadata contains a link between the ORKG Paper's DOI and the Crossref DOI of the original article. Other PIDs (for example, contributor ORCID IDs and organization IDs) are also specified in the metadata. This rich and interlinked metadata is shared with DataCite, which in turn shares it with other scholarly communication infrastructures (e.g., Crossref, OpenAIRE, ORCID). We ensure that ORKG Papers are discoverable in global scholarly infrastructures by exploiting this existing mechanism.

Table 1 presents the metadata shared when registering the DOI of an ORKG Paper with DataCite and also illustrates the relationship between the DOI of the ORKG Paper and the DOI of the original article.
The main metadata elements are:

- *identifier*: The DOI assigned to the ORKG Paper.
- *creators*: The creator's name and ORCID. *creatorName* specifies the name whereas *nameIdentifier* specifies ORCID as the identifier type. If the ORCID is included while persistently identifying the ORKG Paper, DataCite exchanges this metadata with ORCID to acknowledge the contribution in the respective ORCID Record.

---

[2] https://support.datacite.org/docs

[3] https://schema.datacite.org

- *subject*: The research area of the article, e.g. ecology and evolutionary biology.
- *version*: Represents the Paper version.
- *resourceType*: We consider ORKG Papers to be of type Dataset. *relatedIdentifiers*: Links the ORKG Paper entry with the DOI of the original article, whereas *relationType* defines the relationship with the related resource. In our case, the described article is *referenced* by the ORKG Paper.

Table 1. Metadata shared with DataCite to persistently identify ORKG Paper (machine actionable scholarly knowledge).

```xml
<?xml version="1.0" encoding="UTF-8"?>
<resource xmlns="http://datacite.org/schema/kernel-4"
xmlns:xsi="http://www.w3.org/2001/XMLSchema-instance"
xsi:schemaLocation="http://datacite.org/schema/kernel-4
http://schema.datacite.org/meta/kernel-4.3/metadata.xsd">

<identifier identifierType="DOI"> 10.48366/R57590 </identifier>

<titles>
  <title xml:lang="en"> The invertebrate fauna on broom, Cytisus scoparius, in two native
and two exotic habitats [ORKG] </title>
</titles>

<publisher xml:lang="en"> Open Research Knowledge Graph </publisher>

<version> V0.1 </version>

<resourceType resourceTypeGeneral="Dataset"> Paper </resourceType>

<creators>
  <creator>
    <creatorName nameType="Personal"> Heidari, Golsa </creatorName>
    <nameIdentifier schemeURI="http://orcid.org/" nameIdentifierScheme="ORCID">
0000-0002-5071-1658</nameIdentifier>
  </creator>
</creators>

<subjects> <subject xml:lang="en"> Ecology and Evolutionary Biology </subject>
</subjects>

<relatedIdentifiers>
<relatedIdentifier relationType="References" relatedIdentifierType="DOI">
10.1016/S1146-609X(00)00124-7
 </relatedIdentifier>
</relatedIdentifiers>

<descriptions>
 <description descriptionType="Abstract">
```

> The machine-actionable description of an article: The invertebrate fauna on broom, Cytisus
> scoparius, in two native and two exotic habitats which addresses the research problem Testing
> the enemy release hypothesis in invasion biology.
> </description>
> </descriptions>
> </resource>

*Versioning of Published Papers*

ORKG Papers are stored in a Neo4j database and is accessible via REST API. Data in Neo4j
can be edited by anyone. While persistently identifying ORKG Papers, it is ensured that the
state of the published Paper is saved and can no longer be edited. Updates are possible by
creating a new version of the ORKG Paper. The persistently identified versions are stored in
the ORKG versioning service, which uses a PostgreSQL in its backend. Thus, ORKG Papers
are stored in JSON format in PostgreSQL. All changes to ORKG Papers are recorded in Neo4j
and the updated version can be persistently identified with the provenance link to previous
version, thus showing what has been updated across the different versions.

**Discussion**

Based on user stories, we discuss the advantages of persistently identifying ORKG Papers.

- *Credit*: A researcher who publishes an ORKG Paper wants to be credited for her work.
  If she includes her ORCID, DataCite will automatically ensure that this contribution
  is credited on her ORCID Record.
- *Discovery*: A researcher reads an article and discovers that essential information
  contained in the article is described as an ORKG Paper. Such discovery is enabled by
  linking persistent identifiers and automated sharing of metadata. The researcher may
  use DataCite Commons, the PID Graph or discover such relationships directly on the
  article's landing page by the publisher. For instance, by searching the DOI
  *10.1016/S1146-609X(00)00124-7* on DataCite Common[4], the ORKG Paper describing
  information contained in this article can be discovered in the citations section of the
  DataCite Commons page. As shown in Table 2 for the example DOI, a software agent
  can obtain the same information in machine readable form by querying the PID Graph.

---

[4] https://commons.datacite.org/doi.org/10.1016/S1146-609X(00)00124-7

Table 2. Query executed on DataCite PID Graph for the work with DOI name *10.1016/S1146-609X(00)00124-7* to retrieve its citations. The results include the ORKG Paper references by the work.

```
Query:
{ work( id: "10.1016/S1146-609X(00)00124-7")
  {
    titles { title }
doi     citations
{
totalCount
    nodes {
id
      creators { id name }
      titles { title }
} } } }

Result (shortened):
{ "data": {
  "work": {
    "titles": [
{
      "title": "The invertebrate fauna on broom, Cytisus scoparius,in two native and two
exotic habitats"
 }],
   "doi": "https://doi.org/10.1016/S1146-609X(00)00124-7",
   "citations": {"totalCount": 1,
    "nodes": [{ "id": "https://doi.org/10.48366/r57590",
      "creators": [{
        "id": 0000-0002-5398-7086,
        "name": "Heidari, Golsa"
       }],
      "titles": [{
        "title": "The invertebrate fauna on broom, Cytisus scoparius, in two native and two
exotic habitats [ORKG]"
 } ] ] ] } } } }
```

We suggest that PIDs play an important role to persistently identify machine actionable representations of essential information published in articles. The proposed implementation leverages ORKG and DataCite services. Publishing ORKG Papers with DataCite enables their findability in global scholarly communication infrastructures. ORKG also provides functionality to update persistently identified Papers as separate stable versions to allow for evolving content.

The persistent identification of ORKG Papers could make discovering, understanding, and processing (e.g., comparisons) the main research contributions of some research work more efficient, particularly for machines.

Since ORKG supports adding research data and software, for future work, we aim to mine software packages from different data repositories (zenodo, figshare etc.), analyze their metadata and add them in ORKG, as well as link them with their respective Papers. We also aim to perform static code analysis of software packages to determine the datasets used as an input and produced as an output. Ingesting such information in ORKG will determine the relations between software codes, used datasets and linked Papers at large-scale. Thus, users will have quick access to the interlinked scholarly knowledge (Papers, datasets and software).

**Conclusion**

We presented an approach for persistent identification of FAIR scholarly knowledge to ensure its findability in global scholarly communication infrastructures. Here, scholarly knowledge is the essential information published in scholarly articles. The main focus of our work is to make the machine-actionable scholarly knowledge in ORKG (Papers, Comparisons, Reviews (Oelen et al., 2021)) persistently refereanceable, citable and findable in global scholarly communication infrastructures. As such, our work is an important towards FAIR scholarly knowledge, i.e. the FAIRification of this content. The persistent identification of scholarly artefacts should be practiced widely to enable the automatic metadata-based linking of contextual artefacts including datasets, software, instruments, scholarly articles, and workflows. These metadata-based linked artefacts can be discovered across different scholarly infrastructures, thus making access to scholarly knowledge more effective and efficient.